\documentclass[a4paper]{amsart}
\usepackage{amsmath,amssymb, amscd}
\usepackage[urlcolor=blue,breaklinks=true,colorlinks=true,citecolor=red]{hyperref}
\usepackage{pdfsync}
\usepackage{graphicx,a4wide}
\usepackage{cite}
\usepackage{url}

\newtheorem{theorem}{Theorem}[section]

\newtheorem{definition}[theorem]{Definition}
\newtheorem{remark}[theorem]{Remark}

\newcommand{\I}{\mathrm{i}}

\newcommand{\name}[1]{\texttt{#1}}

\begin{document}

\title{Efficient computation of multidimensional theta functions}

\author[J.~Frauendiener]{J\"org Frauendiener}
\address[J.~Frauendiener]{Department of Mathematics and Statistics, University of Otago,     
P.O. Box 56, Dunedin 9054, New Zealand}
\email{joergf@maths.otago.ac.nz}
\author[C.~Jaber]{Carine Jaber}
\address[C.~Jaber]{Institut de 
Math\'ematiques de Bourgogne,
Universit\'e de Bourgogne,
9 avenue Alain Savary,
BP 47970, 21078 Dijon Cedex,
France}
\email{carine.jaber@u-bourgogne.fr}
\author[C.~Klein]{Christian Klein}
\address[C.~Klein]{Institut de 
Math\'ematiques de Bourgogne,
Universit\'e de Bourgogne,
9 avenue Alain Savary,
BP 47970, 21078 Dijon Cedex,
France}
\email{christian.klein@u-bourgogne.fr}

\begin{abstract}
An important step in the efficient computation of multi-dimensional 
theta functions is the construction of appropriate symplectic 
transformations for 
a given Riemann matrix assuring a rapid convergence of the theta series. An 
algorithm is presented to approximately map the Riemann matrix 
to the Siegel fundamental domain. The shortest vector of the lattice 
generated by the Riemann matrix is identified exactly, and the algorithm 
ensures that its length is larger than $\sqrt{3}/2$. The approach is 
based on a previous algorithm by Deconinck et al. using the LLL algorithm for 
lattice reductions. Here, the LLL algorithm is replaced by exact 
Minkowski reductions for small genus and an exact identification of 
the shortest lattice vector for larger values of the genus.

\end{abstract}
\maketitle

\section{Introduction}

Multidimensional theta functions are important in many fields of mathematics and
in applications, for instance in the theory of integrable partial differential
equations (PDEs), see e.g.~\cite{algebro}, in conformal field theories and
cryptography. Since they can be seen as the main building block of meromorphic
functions on Riemann surfaces, they appear naturally where Riemann surfaces and
algebraic curves are of importance. They are conveniently defined as a
multi-dimensional series, 
\begin{equation}\label{theta}
    \Theta_{\mathrm{p}\mathrm{q}}(\mathrm{z},\mathbb{B})=
    \sum\limits_{\mathrm{N}\in\mathbb{Z}^g}\exp\left\{
    \I\pi\left\langle\mathbb{B}\left(\mathrm{N}+\mathrm{p}\right),
    \mathrm{N}+\mathrm{p}
    \right\rangle+2\pi \I
    \left\langle \mathrm{z}+\mathrm{q},\mathrm{N}+\mathrm{p}
    \right\rangle\right\}
    \;,
\end{equation}
with $\mathrm{z}\in\mathbb{C}^g$ and the \emph{characteristics} $\mathrm{p}$,
$\mathrm{q}\in{ \mathbb{R}}^g$, where $\left \langle\cdot,\cdot\right\rangle$
denotes the Euclidean scalar product
$\left\langle \mathrm{N},\mathrm{z}\right\rangle=\sum_{i=1}^gN_iz_i$. The matrix
$\mathbb{B}=X+\mathrm{i}Y$ is a \emph{Riemann matrix}, i.e., it is symmetric and
has a positive definite imaginary part $Y$. The latter property ensures that
the series (\ref{theta}) converges uniformly for all $\mathrm{z}$, and that the
theta function with characteristics is an entire function of
$\mathrm{z} \in \mathbb{C}^{g}$.
    
The goal of the present article is the description of an efficient treatment of
theta functions appearing in the context of integrable PDEs such as the
Korteweg-de Vries (KdV) and nonlinear Schr\"odinger (NLS) equations.
Quasiperiodic solutions to these PDEs were given at the beginning of the 1970s
by Novikov, Dubrovin, Its, Matveev, van Moerbeke, Krichever and others in terms
of such theta functions on compact Riemann surfaces, see \cite{algebro,Dintro}
and references therein for a historic account. To study such solutions
numerically, theta functions of the form (\ref{theta}) have to be evaluated for
a given Riemann matrix and characteristics for many values of the argument
$\mathrm{z}\in \mathbb{C}^{g}$. To this end, efficient numerical tools to
evaluate theta functions are needed. The \name{algcurves} package distributed
with Maple originally due to Deconinck and van Hoeij, see
\cite{deco01,RSbookdp}, has integrated theta functions as outlined in
\cite{deconinck03} starting from Maple 7. Note that the \name{algcurves} package is
currently being transferred to a new platform \cite{sage}. A purely numerical
approach to compact Riemann surfaces via algebraic curves was given in
\cite{prd,cam,lmp,alg1,hyper,rsart}, see \cite{KK1,KK2} for 
applications to integrable PDEs. For a review of the current state of the art of
computational approaches to Riemann surfaces, the reader is referred to
\cite{Bob}.

The basic idea of the algorithms for theta functions in \cite{Bob} is to
approximate the expression (\ref{theta}) via a truncated series\footnote{For
  alternative approaches based on arithmetic-geometric means and Newton
  iterations, see \cite{dupond,labrande}, which were developed for studying
  modular functions, i.e., the dependence of theta functions with $\mathrm{z}=0$
  on the Riemann matrix; in practice these approaches appear to be mainly
  interesting if precisions of several thousand bits are needed.}. Obviously the
convergence of the series (\ref{theta}) depends on the bilinear term, more
precisely on the \emph{shortest vector} $\mathrm{N}_{min}$ of the lattice
$\mathbb{Z}^g$ equipped with the inner product defined by the imaginary part $Y$
of the Riemann matrix $\mathbb{B}$:
$\left\langle \mathrm{N}, \mathrm{M} \right\rangle_Y:=\left\langle Y \mathrm{N},
  \mathrm{M} \right\rangle$, $\mathrm{N},\mathrm{M}\in \mathbb{Z}^{g}$. For a given Riemann matrix the shortest vector
$\mathrm{N}_{min}$ is then defined in terms of its squared length
\begin{equation}
   y_{min}=\left\langle \mathrm{N}_{min},
    \mathrm{N}_{min}
    \right\rangle_Y:= \mbox{min}_{\mathrm{N}\in\mathbb{Z}^{g}/\{0\}}\left\langle Y\mathrm{N},
    \mathrm{N}
    \right\rangle
    \label{sv}.
\end{equation}
Obviously, the longer the shortest vector, the more rapid the convergence of the
theta series. Changing the shortest vector can be achieved by changing the
homology basis of the underlying Riemann surface which yields a different but
symplectically equivalent Riemann matrix. This can be achieved by using modular
transformations, i.e., symplectic transformations with integer coefficients to
generate larger norms of the shortest vector in order to accelerate the
convergence of a theta series for given $\mathbb{B}$. Since the behavior of
theta functions under modular transformations is explicitly known, such
transformations can dramatically increase the rate of convergence which is
especially important for larger values of $g$. This approach was for the first
time implemented in an algorithm by Deconinck et.~al.\ in~\cite{deconinck03}.

The main task in this context is the identification of the shortest vector in a
given $g$-dimensional lattice known as the shortest vector problem (SVP). Currently,
there is no known algorithm that would solve this problem in polynomial
time. The LLL algorithm~\cite{LLL} yields an approximation to the shortest
vector in polynomial time but with an error growing exponentially with the
dimension $g$ (though in practice slowly with $g$ such that it can be used for
small genus as an approximation). For these reason, in~\cite{deconinck03} the
SVP was solved approximately via the LLL algorithm. However, since we are
interested in an evaluation of theta functions in a large number of points, it
can be beneficial to identify the shortest vector exactly even for small $g$.
Though it is computationally demanding this knowledge will
accelerate the ensuing evaluation of the theta function (\ref{sv}). Therefore,
we replace the LLL algorithm in~\cite{deconinck03} with an exact Minkowski
reduction for $g\leq 5$, and with an exact solution to the SVP for higher genus.

The paper is organized as follows: in section 2 we summarize mathematical facts
about symplectic transformations, theta functions, and Siegel's fundamental
domain. In section 3 we review various notions of lattice reductions and discuss
the algorithms used for the LLL and Minkowski reduction. In section 4 we present
Siegel's algorithm to ensure that the imaginary part of the transformed Riemann
matrix has a shortest lattice vector of squared length greater than $\sqrt{3}/2$
and discuss examples. We add some concluding remarks in section 5.

\section{Theta functions and symplectic transformations}%
\label{9}
In this section we summarize important properties of multi-dimensional 
theta functions and symplectic transformations. In particular we are 
interested in the behavior of theta functions under symplectic 
transformations and in the Siegel fundamental domain.

\subsection{Symplectic transformations} A Riemann matrix
$\mathbb{B}=X+\mathrm{i}Y$ lies in $\mathbb{H}^{g}$, the Siegel space of
symmetric $g\times g$ matrices with positive definite imaginary part 
$Y$ and real part $X$.
Riemann matrices are only unique up to modular transformations, i.e.,
transformations $\mathcal{A}_{g}\in Sp(2g,\mathbb{Z})$, the symplectic or modular group,
of the form 
\begin{equation}
    \mathcal{A}_{g} = 
    \begin{pmatrix}
        A & B \\
        C & D
    \end{pmatrix}
    \label{siegel1},
\end{equation}
where $A,B,C,D$ are $g\times g$ integer matrices satisfying
\begin{equation}
    \begin{pmatrix}
        A & B \\
        C & D
    \end{pmatrix}^{T} \begin{pmatrix}
        0_{g} & I_{g} \\
        -I_{g} & 0_{g}
    \end{pmatrix}
    \begin{pmatrix}
        A & B \\
        C & D
    \end{pmatrix}
    = \begin{pmatrix}
        0_{g} & I_{g} \\
        -I_{g} & 0_{g}
    \end{pmatrix}
    \label{siegel2};
\end{equation}
here $0_{g}$ and $I_{g}$ are the $g\times g$ null and identity matrix 
respectively. The Riemann matrix transforms under these modular 
transformations $\mathcal{A}_{g}$ (\ref{siegel1}) as 
\begin{equation}
    \mathbb{H}^{g}\mapsto \mathbb{H}^{g}:\quad \mathbb{B}\mapsto 
    \tilde{\mathbb{B}}=(A\mathbb{B}+B)(C\mathbb{B}+D)^{-1}
    \label{siegel3}.
\end{equation}

Siegel~\cite{siegel} gave the following fundamental domain for the modular group in 
which each Riemann surface characterized by its Riemann matrix 
$\mathbb{B}$ corresponds to exactly one point:
\begin{definition}\label{def}
 Siegel's fundamental domain is the subset of $\mathbb{H}^{g}$   such that 
 $\mathbb{B}=X+\mathrm{i}Y\in\mathbb{H}^{g}$ satisfies:
 \begin{enumerate}
     \item  $|X_{nm}|\leq 1/2$, $n,m=1,\ldots,g$,
 
     \item  $Y$ is in the fundamental region of Minkowski reductions,
 
     \item  $|\det(C\mathbb{B}+D)|\geq 1$ for all $C$, $D$  (\ref{siegel2}).
 \end{enumerate}
\end{definition}
Roughly speaking, the three conditions address different parts of the modular
transformation in~(\ref{siegel1}). The first condition in Def.~\ref{def} fixes
the matrices $B$ in (\ref{siegel1}). The second condition refers to Minkowski
reductions~\cite{Min1,Min2} and fixes the unimodular matrices $A$ in
(\ref{siegel1}). Minkowski reductions of the lattice generated by a positive
definite matrix $Y$ are equivalent to the introduction of a reduced lattice
basis, i.e., a collection of vectors of minimal length which can be extended to
a basis of the lattice. This condition will be discussed in more detail in the
following section.  The third condition in Def.~\ref{def} fixes the matrices
$C$, $D$ in (\ref{siegel1}). Since $|\det \mathbb{B}|$ can be viewed as the
`height' of $\mathbb{B}$ (see~\cite{klingen}), this can be seen as a condition
of maximal height.

In genus~1, the above conditions~\ref{def} give the well known 
elliptic fundamental domain,
\begin{equation}
    |X|\leq 1/2,\quad X^{2}+Y^{2}\geq 1
    \label{elliptic}.
\end{equation}
Note that parts of the boundary of the fundamental domain in Def.~\ref{def} have
to be excluded in order to assure that no two different points on the boundary
can be related by a symplectic transformation.
We will not address this problem in this paper since we are mainly interested in
the convergence of the theta series.

Siegel~\cite{siegel} showed that the third of the conditions in Def.~\ref{def}
is equivalent to a finite number of conditions, i.e., just a finite number of
matrices $C$ and $D$ has to be considered. But it is not known how to
efficiently obtain this finite set of matrices. The only case $g>1$ where this
has been achieved appears to be Gottschling's work~\cite{gottschling} for
genus~2. In this case, the fundamental domain is defined by the following set of
inequalities: the standard limits for the real part of the Riemann matrix,
\[|X_{11}|\leq \frac{1}{2},\quad |X_{12}|\leq \frac{1}{2},\quad 
|X_{22}|\leq \frac{1}{2},\]
the Minkowski ordering conditions: 
\begin{equation}
Y_{22}\geq Y_{11}\geq 2Y_{12}\geq 0, \label{mink}
\end{equation}
and the following set of~19 inequalities corresponding to the third 
condition in Def. \ref{def}:
\begin{equation}    
|\mathbb{B}_{11}|\geq 1,\quad |\mathbb{B}_{22}|\geq 1,\quad    
|\mathbb{B}_{11}+\mathbb{B}_{22}-2\mathbb{B}_{12} + \mathbf{e}|\geq1,    
\label{gott1}\end{equation}
where $\mathbf{e}=\pm 1$, and 
\begin{equation}    |\det (\mathbb{B}+S)|\geq 1    
\label{gott2},
\end{equation}
where $S$ are the matrices
\begin{equation}    
\begin{split}    
\begin{pmatrix}        0 & 0  \\        0 & 0    
\end{pmatrix},\quad     
\begin{pmatrix}     \mathbf{e} & 0  \\            0 & 0       
\end{pmatrix},\quad     \begin{pmatrix}     0 & 0  \\       0 & \mathbf{e}    
\end{pmatrix},\quad     \begin{pmatrix}     \mathbf{e} & 0  \\            0 & \mathbf{e}    
\end{pmatrix},\quad     \\    \begin{pmatrix}       \mathbf{e} & 0  \\            0 & -\mathbf{e}   
\end{pmatrix},\quad     \begin{pmatrix}     0 & \mathbf{e}  \\            \mathbf{e} & 0    
\end{pmatrix},\quad     \begin{pmatrix}     \mathbf{e} & \mathbf{e}  \\         \mathbf{e} & 0    
\end{pmatrix},\quad     \begin{pmatrix}     0 & \mathbf{e}  \\            \mathbf{e} & \mathbf{e} 
\end{pmatrix}    \label{gott3} .   
\end{split}
\end{equation}

These conditions are important if modular functions expressed in 
terms of theta functions are studied, see for instance 
\cite{Sarnak,dupond,klkoko,ernstbook} and references therein.

\subsection{Theta functions}
Theta functions  with characteristics   are defined as an infinite  
series (\ref{theta}).  A characteristic is called \emph{singular} if the
corresponding theta function vanishes identically. 
 Of special interest are half-integer 
characteristics with $2\mathrm{p},2\mathrm{q}\in \mathbb{Z}^{g}$.  
Such a 
half-integer characteristic is called \emph{even} if $4\langle 
\mathrm{p},\mathrm{q}\rangle=0\mbox{ mod } 2$ and \emph{odd} 
otherwise. It can be easily shown that 
theta functions with odd (even) characteristic are odd
(even) functions of the argument $\mathrm{z}$.  The theta function with
characteristic is related to the Riemann theta function $\Theta$, the
theta function with zero characteristic $\Theta:= \Theta_{\mathrm{00}}$,
via
\begin{equation}
    \Theta_{\mathrm{pq}}(\mathrm{z},\mathbb{B})=\Theta(\mathrm{z}
    +\mathbb{B}\mathrm{p} + \mathrm{q})\exp\left\{\I\pi
    \left\langle\mathbb{B}\mathrm{p},\mathrm{p}\right\rangle+
    2\pi \I\left\langle\mathrm{p},\mathrm{z} + \mathrm{q}\right\rangle
    \right\}\;.
    \label{thchar}
\end{equation}
From its definition, a theta function has the periodicity properties 
\begin{equation}
    \Theta_{\mathrm{p}\mathrm{q}}(\mathrm{z}+\mathrm{e}_{j}) = 
    \mathrm{e}^{2\pi \mathrm{i}p_{j}}
    \Theta_{\mathrm{p}\mathrm{q}}(\mathrm{z})\;,
    \quad 
    \Theta_{\mathrm{p}\mathrm{q}}(\mathrm{z}+\mathbb{B}
    \mathrm{e}_{j})=
    \mathrm{e}^{-2\pi \I (z_{j}+q_{j}) - \I\pi B_{jj}}
    \Theta_{\mathrm{p}\mathrm{q}}(\mathrm{z})\;
    \label{eq:periodicity},
\end{equation}
where $\mathrm{e}_{j}$ is a vector in $\mathbb{R}^{g}$ consisting of zeros
except for a 1 in j$^\text{th}$ position.  
These periodicity properties (\ref{eq:periodicity}) can
be conveniently used in the computation of the theta function: an arbitrary
vector $\mathrm{z}\in \mathbb{C}^{g}$ can be written in the form
$\mathrm{z}=\hat{\mathrm{z}}+\mathrm{N} + \mathbb{B}\mathrm{M}$ with
$\mathrm{N},\mathrm{M} \in \mathbb{Z}^{g}$, where
$\hat{\mathrm{z}}=\mathbb{B}\hat{\mathrm{p}}+\hat{\mathrm{q}}$ with
$|\hat{p}_{i}|\leq 1/2$, $|\hat{q}_{i}|\leq 1/2$. Thus, it is enough to compute
the theta function for arguments $\hat{\mathrm{z}}$ lying in the fundamental domain of
the Jacobian, i.e., $\mathbb{C}^{g}/\Lambda$, where $\Lambda$ is the period
lattice\footnote{Note, that this lattice $\Lambda$ is not to be 
confused with the lattice generated by the matrix $Y$ discussed in 
the present paper. } formed by $\mathbb{B}$ and the $g$-dimensional identity
matrix, $\hat{\mathrm{z}}=\mathbb{B}\hat{\mathrm{p}}+\hat{\mathrm{q}}$ with
$|\hat{p}_{i}|\leq 1/2$, $|\hat{q}_{i}|\leq 1/2$. For general arguments $\mathrm{z}$ one
computes $\Theta(\hat{\mathrm{z}},\mathbb{B})$ and obtains
$\Theta(\mathrm{z},\mathbb{B})$ from the periodicity
properties~\eqref{eq:periodicity} by multiplying with an appropriate exponential
factor.

To compute the series~(\ref{thchar}), it will be approximated by a 
sum, $|N_{i}|\leq \mathcal{N}_{\epsilon}$, $i=1,\ldots,g$, where the 
constant $\mathcal{N}_{\epsilon}$ is chosen such that all omitted 
terms in (\ref{theta}) are smaller than some prescribed value of 
$\epsilon$. Since we work in double precision, we typically choose 
$\epsilon=10^{-16}$, i.e., of the order of the smallest difference 
between floating point numbers that can be handled in Matlab.  Note 
that in contrast to~\cite{deconinck03}, we do not give a specific 
bound for each $N_{i}$, $i=1,\ldots,g$, i.e., we sum over a 
$g$-dimensional sphere instead of an ellipsoid. The reason for this is that 
it does  not add much to the computational cost, but that it simplifies a 
parallelization of the computation of the theta function in which we 
are interested. Taking into account that we can choose $\mathrm{z}$ in the 
fundamental domain of the Jacobian because of (\ref{eq:periodicity}), 
we get with (\ref{sv}) for the Riemann theta function the estimate
\begin{equation}
    \mathcal{N}_{\epsilon}> \sqrt{-\frac{\ln \epsilon}{\pi 
    y_{min}}}+\frac{1}{2}
    \label{ne}.
\end{equation}
Thus the greater the norm of the shortest lattice vector, the more 
rapid will be the convergence of the theta series.

The action of the modular group on theta functions is known, see for 
instance \cite{algebro,fay,Mum}. One has
\begin{equation}
    \Theta_{\tilde{\mathrm{p}}\tilde{\mathrm{q}}}
(\mathcal{M}^{-1}\mathrm{z},\tilde{\mathbb{B}}) = 
    k\sqrt{\det(\mathcal{M})}\exp\left(\frac{1}{2}\sum_{i\leq 
    j}^{}z_{i}z_{j}\frac{\partial}{\partial 
    \mathbb{B}_{ij}}\ln\det\mathcal{M}\right)\Theta_{\mathrm{p}\mathrm{q}}(\mathrm{z}),
    \label{thetamod}
\end{equation}
where $\tilde{\mathbb{B}}$ is given by (\ref{siegel3}),  where $k$ is 
a constant with respect to $\mathrm{z}$, and where
\begin{equation}
    \mathcal{M}=C\mathbb{B}+D,\quad 
    \begin{pmatrix}
        \tilde{\mathrm{p}} \\
        \tilde{\mathrm{q}}
    \end{pmatrix}=
    \begin{pmatrix}
        D& -C \\
        -B & A
    \end{pmatrix}    \begin{pmatrix}
        \mathrm{p} \\
        \mathrm{q}
    \end{pmatrix}
+\frac{1}{2}  \begin{pmatrix}  \mbox{diag}(CD^{T}) \\
       \mbox{diag} (AB^{T})
    \end{pmatrix}
    \label{thetamod2},
\end{equation}
where diag denotes the diagonal of the matrices $AB^{T}$ and 
$CD^{T}$.

\section{Lattices reductions}
In this section we address the second condition in the definition \ref{def} of
Siegel's fundamental domain, the Minkowski reduction of the imaginary part of
the Riemann matrix. This classical problem is related to Euclidean lattices in
$g$ dimensions and the search for efficient bases for them, i.e., a basis
consisting of shortest possible lattice vectors. Finding such a basis is
known as \emph{lattice reduction}. Below we will summarize basic facts on
lattices and their reductions and give a brief review of approximative and exact
approaches to lattice reductions.

\subsection{Lattices and lattice reductions}

In the context of the present paper, we are concerned with lattices, i.e.,
discrete additive subgroups of $\mathbb{R}^{g}$, of the form 
\begin{equation}
    \mathcal{L}(t_{1},\ldots,t_{g}) = 
    \left\{ TN\bigm| N\in\mathbb{Z}^{g}\right\}
    \label{lattice},
\end{equation}
where $T = [t_1,t_2,\ldots,t_g] \in \mathbb{R}^{g\times g}$ has rank $g$. Thus,
the lattice consists of all linear combinations with integer coefficients of the
$g$ linearly independent vectors $t_{i}$, the columns of $T$. The vectors
$t_{i}$ form the \emph{lattice basis}. Different lattice bases are related via
unimodular transformations, $\tilde{T}=TA$ where $A$ is an integer matrix with
$|\det(A)|=1$. This implies in particular that $\det(T^{T}T)$ is an invariant of
such unimodular transformations. The length of a lattice vector $t_{i}$,
$i=1,\ldots,g$, is given by its Euclidean norm
$||t_{i}||^{2}=\sum_{j=1}^{g}T_{ji}^{2}$.

In the context of lattices there are two equivalent points of view. The first
one, which we took above, is to consider the lattice as being generated by a
basis $T$, and then define the matrix $Y=T^TT$ of inner products of the basis
vectors. The complementary point of view is to take a lattice always as being
represented as $\mathbb{Z}^g$ but equipped with a positive definite bilinear
form $Y$ defining the lengths and angles of the lattice vectors. The two view
points are both useful and we can switch between them quite easily: given the
matrix $T$ of basis vectors the bilinear form is represented by $Y=T^TT$, and
given a symmetric positive definite matrix $Y$ we can obtain a matrix $T$
representing a lattice basis by a Cholesky decomposition, which yields an upper
triangular matrix $R$ with $Y=R^TR$.


It is well known that in $\mathbb{R}^{g}$ it is always possible to introduce an
orthonormal basis, and the Gram-Schmidt procedure allows to determine such a
basis from a given general one. In the discrete case of a lattice, in general
there will be no basis consisting of orthogonal vectors, and there is a lower
bound on the length of vectors in a lattice. A basis is considered
\emph{reduced} if it satisfies certain conditions. In general, the goal is to
find a basis of vectors of minimal length with a minimal deviation from
orthogonality. Thus, important issues in lattice theory are the shortest vector
problem (SVP), i.e., the determination of the shortest non-zero lattice vector,
and the closest vector problem (CVP), i.e., the location of the lattice vector
closest to a given point $\mathbf{x} \in \mathbb{R}^g$, see for
instance~\cite{stehle} for a recent review.

The strongest known lattice reduction is \emph{Minkowski reduction}: a basis
for a lattice $\mathcal{L}$ generated by a matrix $Y$ is
Minkowski reduced if it consists of shortest lattice vectors 
which can be extended to a basis of $\mathcal{L}$. 

In a more
narrow sense, a symmetric and positive definite matrix is Minkowski reduced if it satisfies 
Minkowski's conditions,
\begin{equation}
    ||x_{1}t_{1}+\ldots+x_{g}t_{g}||\geq ||t_{i}||,
    \label{minkcond}
\end{equation}
for all $1\leq i \leq g$ and for all integers $x_{1},\ldots,x_{g}$ 
such that 
$\mbox{gcd}(x_{1},\ldots,x_{g})=1$. A minimal set of these conditions was
given by Minkowski \cite{Min1,Min2} for
$g\leq4$. For $g=5,6,7$ these conditions were presented in \cite{tam2}. Note that the number
of Minkowski reduction conditions grows rapidly with the dimension of the
lattice, for $g=7$ there are 90 000 conditions in \cite{tam2}. The corresponding
conditions do not appear to be known for $g>7$. 

The \emph{simple Minkowski
  reduction} is achieved by the additional condition $Y_{i,i+1}>0$ for
$i=1,\ldots,g-1$, which fixes the orientation of the vectors. 

\begin{remark}
    For $g=2$, the Minkowski fundamental domain (condition 2 of 
    Def.~\ref{def}), i.e., the 
    fundamental domain of the unimodular group, 
    is given by the simple Minkowski reduction. But
    for $g>2$ the simple Minkowski reduction does not define 
    the Minkowski fundamental domain, see \cite{tammela}. 
    The fundamental domain for $g=3$ is given in \cite{tammela}, but 
    the corresponding conditions in higher dimensions appear to be 
    unknown. 
\end{remark}

The condition that the set of shortest lattice vectors have to form a basis of
the lattice is problematic from an algorithmic point of view, in addition the
Minkowski conditions are not known for $g>7$. Therefore the lattice reduction by
Hermite \cite{hermite}, Khorkine and Zolotareff \cite{KZ} (HKZ) is generally
preferred in applications for $g\geq 7$: in this reduction, the shortest lattice
vector is identified and a unimodular transformation is found such that this
vector is used as the first basis vector. This means, it appears as
$\tilde{T}_{i1}$, the first column in the transformed matrix $\tilde{T}$; next, the shortest vector of the
$(g-1)\times(g-1)$ dimensional matrix $\tilde{T}_{i,j}$, $i,j=2,\ldots,g$ is
identified, a unimodular transformation to put this vector as the first of the
transformed $(g-1)\times(g-1)$ dimensional matrix; then an SVP is solved for the
$(g-2)\times(g-2)$ dimensional matrix obtained after taking off this vector and
the first line of the matrix and so on.

Since both Minkowski and HKZ reduction require the solution of SVPs for which no
algorithms in polynomial time are known, see \cite{stehle}, the same applies to
these reductions. Therefore, the LLL algorithm \cite{LLL} is often used since it
converges in polynomial time. It essentially applies Gauss' algorithm 
\cite{lagrange,gauss} in
dimension 2 to pairs of vectors in higher dimensions, see below. The problem
with the LLL algorithm is that it approaches the solution of the SVP with an
error that grows exponentially with the dimension of the lattice, see the
examples below and in the next section. Note that all above mentioned
reductions, Minkowski, HKZ and LLL lead to the same result in dimension $g=2$
where they agree with the result of the Gauss algorithm: the two shortest
lattice vectors are identified there.

\subsection{Algorithms}
An important step in most lattice reduction algorithms is the SVP 
or more generally, the CVP. One distinguishes exact 
algorithms to find the respective vectors, which typically find the 
vectors by more or less sophisticated enumeration of all possible 
vectors, approximative algorithms, or probabilistic algorithms of 
Monte-Carlo type, see \cite{stehle} for a review. It turns out that 
for $g\leq 40$, enumerative algorithms can be more efficient than 
probabilistic ones. Since we are mainly interested in the case of 
small genus ($g\leq 20$), we concentrate here on the former. As 
before, we always put $Y=T^{T}T$ where without loss of generality we may assume
that $T$ is an upper triangular matrix formed by vectors $t_{1},\ldots,t_{g}$. 

\paragraph{\textbf{Gauss reduction}}
Gauss reduction provides an algorithm to find the Minkowski and HKZ reduced form
of a two-dimensional lattice formed by two vectors $t_{1}$ and $t_{2}$. The
algorithm identifies the two shortest vectors in this lattice. Size reductions
motivated by Gram-Schmidt type formulae and swapping of vectors are alternated
in this algorithm until it terminates. A basis is \emph{size reduced} if condition
(\ref{mink}) is satisfied.

\begin{itemize}
    \item  In the size reduction step, one puts 
    \begin{equation}
        \tilde{t}_{2}= t_{2}-[\mu+1/2]t_{1},\quad \mu=\frac{\langle 
	t_{1},t_{2}\rangle}{||t_{1}||^{2}}
        \label{gauss1},
    \end{equation}
    i.e., a linear combination of the two vectors with the rounded 
    Gram-Schmidt factor~$\mu$. 

    \item  If the resulting vector $\tilde{t}_{2}$ is shorter than 
    $t_{1}$, the vectors are swapped, and a further size reduction 
    step follows. The algorithm terminates when the vector 
    $\tilde{t}_{2}$ of the size reduction step (\ref{gauss1}) is 
    longer than $t_{1}$.
\end{itemize}

\paragraph{\textbf{LLL reduction}}
The LLL algorithm essentially generalizes Gauss' algorithm to higher 
dimensions than~2. A parameter $\delta$ is chosen such that 
$1/4<\delta\leq 1$ (the algorithm is not polynomial in time for 
$\delta=1$). The Gram-Schmidt vectors $t^{*}_{k}$ and  matrix 
$\mu_{i,k}$, $i,k=1,\ldots,g$ given by 
$$t_{i}^{*}=t_{i}-\sum_{j=1}^{i-1}\mu_{i,j}t^{*}_{j},\quad 
\mu_{i,k}=\frac{\langle t_{i},t_{k}^{*}\rangle}{||t_{i}^{*}||^{2}}$$
are computed. If 
the LLL condition
\begin{equation}
    ||t_{k}^{*}||^{2}\geq (\delta-\mu_{k,k-1}^{2})||t_{k-1}^{*}||^{2}
    \label{LLLcond}
\end{equation}
is not satisfied for some $k$ (starting with $k=2$), the reduction step of the
Gauss algorithm and a possible swap of the vectors are applied to this pair of
vectors. The Gram-Schmidt matrix is updated, and if $t_{k-1}^{*}$ has
changed, the algorithm continues with $k$ replaced by $k-1$. Otherwise one
passes to the pair of vectors $t^{*}_{k+1}$, $t^{*}_{k}$. Reductions and swaps of
pairs of vectors are continued until the LLL condition (\ref{LLLcond}) is
satisfied for all $k=1,\ldots,g$.

\paragraph{\textbf{Sphere decoding}}
The basic idea of sphere decoding algorithms is that for the CVP for a general
point $\mathrm{x}\in \mathbb{R}^g$, all lattice points $\mathrm{z}$ 
inside a sphere of radius $\rho$ centered 
at $\mathrm{x}$ are enumerated (for treating the SVP, one takes 
$\mathrm{x}=0$ excluding the
lattice point $\mathrm{z}=0$ as a possible solution of the problem). 
\begin{itemize}
\item To obtain an estimate for the radius $\rho$, we choose the norm of the
  first vector. This is not the optimal choice for $\rho$, see for instance
  \cite{babai}, but it guarantees that the algorithm will always find a vector
  with length $\rho$ or shorter.
\item The problem is then to find lattice vectors $\mathrm{z}$ satisfying 
$||T\mathrm{z}-\mathrm{x}||<\rho$. Since $T$ is upper triangular this inequality becomes
\begin{equation}
(T_{gg}z_{g}-x_{g})^{2} + (T_{g-1,g-1}z_{g-1} + T_{g-1,g}z_{g} - x_{g-1})^{2}+
\ldots<\rho^{2}.   
    \label{rec}
\end{equation}
In particular this implies 
\begin{equation}
|z_{g}| < |x_g \pm \rho|/T_{gg}.\label{eq:z_ineq}
\end{equation}
Equation~(\ref{rec}) suggests a recursive implementation of the algorithm, see
\cite{pohst}: the integers within the above limits are enumerated as possible
candidates for $z_g$; for each such possible component $z_g$ a CVP in
dimension~$g$ is equivalent to a CVP in dimension $g-1$. Thus, a CVP in dimension
$g$ can be reduced to a finite number (corresponding to the possible choices for
$z_{g}$) of CVPs in dimension $g-1$. The search process starts at level~$g$ and
goes down recursively to level 1 to solve a one-dimensional problem. 
\item 
  On each level, the enumeration of the integer candidates for the component
  $z_g$ which are restricted by~\eqref{eq:z_ineq} uses the strategy by Schnorr and Euchner \cite{SE}, i.e., one starts with 
$z_g = n_0:= [x_g/T_{gg}]$ and then in a zig-zag approach $z_g = n_0\pm1,n_0\pm2,\ldots$ are explored until the limits of~\eqref{eq:z_ineq} are reached.
\item All vectors with a length larger than $\rho$ are rejected. If a vector
  with a length smaller than $\rho$ is found, the radius $\rho$ is updated with
  this smaller length and the procedure is continued with the new value.  This
  is done until the shortest nontrivial vector $t_{min}$ is identified.

\item When used in the context of Minkowski reduction, only vectors 
with mutually prime entries are considered.

\item A unimodular matrix $\mathcal{M}$ is constructed such that this vector 
is the first in the matrix $T\mathcal{M}$, i.e., 
$\mathcal{M}^{-1}t_{min}=e_{1}$, where $e_{1}=(1,0,\ldots,0)^{T}$. 

\end{itemize}

\paragraph{\textbf{Minkowski reduction}}
The idea of the algorithm \cite{ZQW} for a Minkowski reduction is to apply the
above SVP algorithm successively to a lattice in order to find a set of shortest
lattice vectors. In the first step, an SVP is solved and a unimodular matrix is
identified such that this vector appears as the first vector of the transformed
matrix as above. At the $p^{\text{th}}$ step, we have a basis
$B_{p} = \{\tilde{t}_{1},\ldots,\tilde{t}_{p-1},t_{p},\ldots,t_{g}\}$. To extend
this basis to a Minkowski reduced basis, the $p^{\text{th}}$ reduced basis
vector must satisfy:
\[
||\tilde{t}_{p}||=\mbox{min}\{||T\mathrm{z}||: \mathrm{z}\in\mathbb{Z}^{g}, 
\mbox{gcd}(z_{p},\ldots,z_{g})=1\}.
\]
The gcd condition is directly implemented in the SVP algorithm. 
To extend $\{\tilde{t}_{1},\ldots,\tilde{t}_{p}\}$ to a basis, one has to find a 
unimodular matrix $Z$ such that $B_{p+1}=B_{p}Z$, i.e., a unimodular 
matrix which does not affect the first $p-1$ vectors, and which has 
$\mathrm{z}$  in $p$th position, where $\tilde{t}_{p} = B_{p}\mathrm{z}$. 

For $1\leq i\leq g$, the $i$-th Minkowski's successive minimum is 
defined as the 
radius of the smallest closed ball centered at the origin containing 
at least $i$ linearly independent lattice vectors.   Note that 
Minkowski showed that in dimensions $g>4$, the vectors realizing 
Minkowski's successive minima may not form a lattice basis. Thus a 
Minkowski reduction algorithm based on SVPs only can fail. The construction 
of an appropriate unimodular matrix 
is thus crucial, see also \cite{helfrich}. Note that the corresponding 
reduced matrix will in general not satisfy the Minkowski reduction 
conditions which are not even explicitly known for $g>7$. 
Since we are here mainly interested in the 
shortest lattice vector, we do not explore Minkowksi reductions for 
$g>5$.

\subsection{Examples}
To illustrate the difference between an LLL and a Minkowski reduced 
basis, we consider a symmetric real $4\times 4$ matrix (for the ease 
of representation, we only give 4 digits throughout the paper)
\begin{verbatim}
    Y =

    0.7563    0.4850    0.4806    0.3846
    0.4850    1.3631    0.2669   -0.3084
    0.4806    0.2669    0.7784   -0.4523
    0.3846   -0.3084   -0.4523    1.7538.
\end{verbatim}
This matrix was created using a $4\times4$ matrix $L$ with random 
entries, and then setting $Y=L^{T}L$. As usual we put $Y=T^{T}T$ where 
the upper triangular matrix $T$ is obtained from $Y$ via a Cholesky 
decomposition. 

To this matrix $T$, we apply the algorithm for Minkowski reductions 
discussed above based on a successive finding of shortest lattice 
vectors. The found matrix $\tilde{Y}=\tilde{T}^{T}\tilde{T}$ is then 
postprocessed to ensure the simple
Minkowski reduction condition $\tilde{Y}_{i,i+1}\geq0$, $i=1,2,3$, i.e., a 
unimodular matrix $\tilde{Z}$ is constructed such that 
 $\tilde{Z}^T\tilde{Y}\tilde{Z}$ has positive elements in the right parallel to 
the diagonal. This leads to the matrix
\begin{verbatim}
    0.5321    0.2058   -0.1639    0.0181
    0.2058    0.5735    0.0920    0.2634
   -0.1639    0.0920    0.5741    0.1364
    0.0181    0.2634    0.1364    0.6535.
\end{verbatim}
In particular it can be seen that the squared length of the shortest lattice 
vector is $0.5321$, the $(11)$ element of the matrix.

An LLL reduction with $\delta=3/4$ of the matrix $Y$ leads to 
\begin{verbatim}
    0.7563   -0.2757    0.3182   -0.1089
   -0.2757    0.5735    0.0920    0.2634
    0.3182    0.0920    0.5741    0.1364
   -0.1089    0.2634    0.1364    0.6535.
\end{verbatim}
The length of the shortest vector identified by the LLL algorithm is 
in this example $0.5735$, thus longer than the shortest vector of the 
lattice which is still $0.5321$ since both the LLL and the 
Minkowski reduction of a lattice are obtained via unimodular 
transformations. The latter obviously do not change the length of the 
shortest vector.  Thus, this example shows 
that the LLL algorithm can 
lead to a considerable overestimation of the length of the shortest 
vector even for small size of the matrix. The effect is known to grow 
exponentially with the size of the matrix. 

Note that in the above example, the shortest vector appears as the second vector in 
contrast to a Minkowski ordered matrix where the shortest vector is 
always the first. 
\begin{remark}
    An LLL reduced matrix is always ordered in accordance with the 
    LLL condition (\ref{LLLcond}). Thus there is no reason why the 
    shortest vector should appear in the first position as in 
    Minkowski reduced matrices. This is especially important in the 
    context of the Siegel algorithm to be discussed in the following 
    section, where the shortest vector is always assumed to be the 
    first of the matrix. 
\end{remark}

To illustrate this aspect even more, we consider another example of a 
random matrix,
\begin{verbatim}
    Y =

    1.7472    0.5191    1.0260    0.6713
    0.5191    1.3471    0.2216   -0.5122
    1.0260    0.2216    0.6801    0.4419
    0.6713   -0.5122    0.4419    0.7246.
\end{verbatim}
The Minkowski reduction yields
\begin{verbatim}
    0.2205    0.0443    0.0342    0.0351
    0.0443    0.3636    0.1660   -0.0294
    0.0342    0.1660    0.3688    0.1516
    0.0351   -0.0294    0.1516    0.3753.
\end{verbatim}
The corresponding LLL reduced matrix ($\delta=3/4)$ takes the form
\begin{verbatim}
    0.3753    0.0294   -0.1516    0.0351
    0.0294    0.3636    0.1660   -0.0443
   -0.1516    0.1660    0.3688   -0.0342
    0.0351   -0.0443   -0.0342    0.2205.
\end{verbatim}
In this case the shortest vector is found by the LLL algorithm in 
contrast to the previous example, but it appears as the last vector. 
The first vector has with $0.3753$ almost twice the length of the 
shortest vector, $ 0.2205$.

\section{Approximation to the Siegel fundamental domain}
In this section, we review an algorithm due to Siegel \cite{siegel} to
approximate the Siegel fundamental domain, which has been implemented together
with the LLL algorithm in \cite{deconinck03}. This algorithm is used here
together with an exact determination of the shortest lattice vector. As an
example we consider the Fricke-Macbeath curve~\cite{fricke,macbeath}, a curve of
genus 7 with the maximal number of automorphisms.

\subsection{Siegel's algorithm}
Whereas Siegel's fundamental domain as defined in Def.~\ref{def} is 
an important theoretical concept in symplectic geometry, its 
practical relevance is limited since no constructive approach exists 
to actually identify the domain for $g>2$: the first condition on the 
components of the matrix of the real part $X$ of $\mathbb{B}$ is straight 
forward. But as discussed in the previous section, already the 
Minkowski fundamental domain appearing in the second condition of 
Def.~\ref{def} is only known for $g\leq 3$. The third condition of 
Def.~\ref{def} is, however, the least studied one. Siegel \cite{siegel} showed that 
it is equivalent to a finite number of conditions, but these 
conditions, except for the classical case $g=1$ in (\ref{elliptic}), are only known for 
$g=2$ in the form (\ref{gott1}) and (\ref{gott2}) due to Gottschling \cite{gottschling}.  

However, Siegel \cite{siegel} gave an algorithm to approximately 
reach the fundamental domain. He proved the following 
\begin{theorem}\label{thm}
   Any Riemann matrix $\mathbb{B}=X+\mathrm{i}Y\in \mathbb{H}^{g}$ 
   with real and imaginary part $X$ respectively $Y$
   can be transformed by 
   a symplectic transformation (\ref{siegel1}--\ref{siegel3}) to 
   a Riemann matrix satisfying the following conditions:
   \begin{enumerate}\parskip=0pt
   \item $|X_{nm}|\leq1/2$, for $n,m=1...g$, 
   \item the squared length of the shortest lattice vector of the lattice generated by $Y$
     is greater than or equal to $\sqrt{3}/2$,
   \end{enumerate}
\end{theorem}

The proof in \cite{siegel}, see also \cite{deconinck03}, is 
constructive and leads naturally to an algorithm:
\begin{proof}
  As already mentioned in the previous section, the first condition can be
  always achieved by an appropriate choice of the matrix $B$ in (\ref{siegel1}),
  $B=[X]$, i.e., each component of $B$ is the integer part
  of the corresponding component of $X$.

  For the second condition, we assume that the shortest vector of the lattice
  generated by $T$, where $T$ is the Cholesky decomposition of $Y=T^{T}T$, is
  the first vector of $T$. It is discussed in the previous section that this can
  be always achieved.  Siegel showed that the determinants of the imaginary part
  of two Riemann matrices $\tilde{\mathbb{B}} =\tilde{X}+\mathrm{i}\tilde{Y}$
  and $\mathbb{B}=X+\mathrm{i}Y$ related by a symplectic transformation
  (\ref{siegel3}) satisfy
  \begin{equation}
    |\det(\tilde{Y})| = \frac{|\det(Y)|}{|\det(C\mathbb{B}+D)|^{2}}.
    \label{det}
  \end{equation}
  If one considers the \emph{quasi-inversion}
  \begin{align}
    A&= 
       \begin{pmatrix}
         0 & \mathbf{0}^T_{g-1} \\
         \mathbf{0}_{g-1} & \mathbf{1}_{g-1,g-1}
       \end{pmatrix},\quad B=
                              \begin{pmatrix}
                                -1 & \mathbf{0}^T_{g-1} \\
                                \mathbf{0}_{g-1} & \mathbf{0}_{g-1,g-1}
                              \end{pmatrix},\nonumber \\
    C &=
        \begin{pmatrix}
          1 & \mathbf{0}^T_{g-1} \\
          \mathbf{0}_{g-1} & \mathbf{0}_{g-1,g-1}
        \end{pmatrix},\quad D =
                               \begin{pmatrix}
                                 0 & \mathbf{0}^T_{g-1} \\
                                 \mathbf{0}_{g-1} & \mathbf{1}_{g-1,g-1}
                               \end{pmatrix},       \label{quasi}
  \end{align}
  where $\mathbf{0}_{g-1}$ is the column vector of $g-1$ zeros, equation (\ref{det}) takes the form
  \begin{equation}
    |\det(\tilde{Y})| = \frac{|\det(Y)|}{|\mathbb{B}_{11}|^{2}}.
    \label{det2}
  \end{equation}
  This leads to the following algorithm:
  \begin{enumerate}
  \item choose $A$ in (\ref{siegel1}) such that the shortest lattice vector
    appears as the first vector of $T$;

  \item choose $B$ in (\ref{siegel1}) such that the real part of
    $\hat{\mathbb{B}}=A^{T}\mathbb{B}A$ has components $|\hat{X}_{nm}|\leq 1/2$, for
    $n,m=1,\ldots,g$;

  \item if $|\hat{\mathbb{B}}_{11}| \geq 1$, terminate the algorithm; if not,
    apply the quasi-inversion (\ref{quasi}) and continue with step 1 of the
    algorithm for the resulting $\mathbb{B}$.
  \end{enumerate}
  Because of (\ref{det2}) the modulus of the determinant of the imaginary part
  of the transformed Riemann matrix increases with each application of
  step~(3). Since Siegel \cite{siegel} has shown that there exists only a finite
  number of symplectic transformations leading to increasing $|\det(Y)|$ and
  that this determinant will be eventually greater than or equal to 1, the algorithm
  terminates after a finite number of steps. Then $Y_{11}$ is the squared
  length of the shortest lattice vector by construction. Since we have
  $|\mathbb{B}_{11}|\geq1$, this implies $Y_{11}^{2}+X_{11}^{2}\geq1$. Since
  $|X_{11}|\leq 1/2$, one has $Y_{11} \geq \sqrt{3}/2$. This proves the
  theorem.
\end{proof}
\begin{remark}
  The fact that the squared length of the shortest vector $y_{min}$ of the
  lattice generated by $Y$ is always greater than $\sqrt{3}/2$ implies that a
  general estimate can be given for the cutoff $\mathcal{N}_{\epsilon}$ in
  (\ref{ne}). For an $\epsilon=2.2\times10^{-16}$, the smallest difference between
  two floating point numbers that Matlab can represent, we find
  $\mathcal{N}_{\epsilon}\approx 4.1$. This means that with an
  $\mathcal{N}_{\epsilon}=4$ the neglected terms in the theta series
  (\ref{theta}) will be of the order of $10^{-14}$, the order of the rounding
  errors, and smaller. For longer shortest vectors, even smaller values of
  $\mathcal{N}_{\epsilon}$ are possible.
\end{remark}

Note that the algorithm \cite{deconinck03} implemented in Maple uses 
the LLL algorithm on $Y$ instead of an exact determination of the shortest 
lattice vector. As discussed in the previous section and illustrated 
below, this is considerably more rapid than an exact determination of 
the vector, but can lead to exponentially (in the dimension $g$) 
growing errors in this context. Since the convergence of the theta 
series is directly related to the shortest vector, we opt here for an 
exactly determined shortest vector. 
If the LLL algorithm is applied, the length of the shortest vector is 
only approximately identified (with an error growing exponentially 
with the dimension). Thus the
the cutoff $\mathcal{N}_{\epsilon}$ (\ref{ne}) has to be based on an 
estimate of the length of the shortest vector which is not provided 
by the algorithm.

\begin{remark}
    In this article, we use the cutoff (\ref{ne}) for all 
    lattice vectors appearing in the theta sum (\ref{theta}), because the summation over a $g$-dimensional 
    sphere can be more easily parallelized. If a summation over an 
    ellipsoid as in \cite{deconinck03} is applied, 
    a different cutoff can be used for each lattice 
    vector. In this case a full 
    Minkowski reduction will be beneficial, whereas in our case, the 
    exact determination of the shortest lattice vector is sufficient. 
\end{remark}

\subsection{Example}
As an example we want to study the Riemann matrix of the Fricke-Macbeath surface
\cite{fricke,macbeath}, a surface of genus $g=7$ with the maximal number $84(g-1)=504$ of
automorphisms. It can be defined via the algebraic curve
\begin{equation}
    f(x,y):=1+7yx+21y^2x^2+35x^3y^3+28x^4y^4+2x^7+2y^7=0.
    \label{FM}
\end{equation}
The code \cite{rsart} produces for this curve the following Riemann 
matrix\footnote{For the ease of the reader, we present only 4 digits 
though the Riemann matrix is computed with an error of the order of 
$10^{-10}$.}
\begin{verbatim}
RieMat =

  Columns 1 through 4

   1.0409 + 1.3005i   0.0530 + 0.3624i   0.3484 + 0.0000i   0.2077 + 0.6759i
   0.0530 + 0.3624i  -0.5636 + 1.0753i   0.0187 - 0.5975i   0.6749 + 0.3001i
   0.3484 + 0.0000i   0.0187 - 0.5975i   1.0544 + 1.7911i   0.3220 - 1.0297i
   0.2077 + 0.6759i   0.6749 + 0.3001i   0.3220 - 1.0297i  -0.0978 + 1.7041i
  -0.2091 - 0.2873i   0.1220 - 0.5274i   0.3029 + 0.8379i  -0.7329 - 0.8055i
  -0.1064 - 0.4257i   0.1205 - 0.1783i  -0.2297 - 0.3668i  -0.0714 - 0.1766i
   0.3590 + 0.5023i   0.1990 - 0.1118i   0.3495 - 0.0499i  -0.0415 + 0.5448i

  Columns 5 through 7

  -0.2091 - 0.2873i  -0.1064 - 0.4257i   0.3590 + 0.5023i
   0.1220 - 0.5274i   0.1205 - 0.1783i   0.1990 - 0.1118i
   0.3029 + 0.8379i  -0.2297 - 0.3668i   0.3495 - 0.0499i
  -0.7329 - 0.8055i  -0.0714 - 0.1766i  -0.0415 + 0.5448i
   1.1824 + 1.0163i   0.4425 + 0.2592i   0.0835 - 0.2430i
   0.4425 + 0.2592i   0.2815 + 0.7791i  -0.6316 - 0.0369i
   0.0835 - 0.2430i  -0.6316 - 0.0369i   0.2315 + 0.6895i.
\end{verbatim}

\begin{remark}
    Since we work with finite precision, rounding is an issue also in the 
    context of lattice reductions. The code \cite{rsart} generally 
    produces results with a tolerance Tol between $10^{-10}$ and 
    $10^{-14}$, which appears for instance in the form of an asymmetry of the 
    computed
    Riemann matrix of the order of Tol. Since in lattice reductions 
    the components of the Riemann matrix are multiplied with integers, 
    these errors will be amplified. Thus a rounding 
     of an order of magnitude larger than Tol is 
    necessary in practice.
\end{remark}

After LLL reduction the first basis vector of the lattice is found to have
squared norm $1.3005$
i.e., the (11) component of the imaginary part of the above Riemann 
matrix. Note that the lattice basis is almost LLL reduced, there are 
only minor effects of the LLL algorithm applied to this matrix.  
Since the norm of the shortest vector is greater than $\sqrt{3}/2$, no quasi-inversion is 
applied.
An ensuing shift of the real part leads to the matrix
\begin{verbatim}
W =

  Columns 1 through 4

   0.0409 + 1.3005i   0.0530 + 0.3624i  -0.4849 - 0.6245i  -0.1064 - 0.4257i
   0.0530 + 0.3624i   0.4364 + 1.0753i  -0.3594 - 0.6598i   0.1205 - 0.1783i
  -0.4849 - 0.6245i  -0.3594 - 0.6598i  -0.4706 + 1.3844i  -0.1946 - 0.1178i
  -0.1064 - 0.4257i   0.1205 - 0.1783i  -0.1946 - 0.1178i   0.2815 + 0.7791i
   0.3590 + 0.5023i   0.1990 - 0.1118i  -0.0510 - 0.0073i   0.3684 - 0.0369i
  -0.4511 + 0.1383i  -0.0171 + 0.2485i  -0.0543 - 0.3239i   0.3907 - 0.1531i
   0.2684 - 0.2975i  -0.4161 + 0.2521i   0.0481 + 0.3949i  -0.2437 - 0.3094i

  Columns 5 through 7

   0.3590 + 0.5023i  -0.4511 + 0.1383i   0.2684 - 0.2975i
   0.1990 - 0.1118i  -0.0171 + 0.2485i  -0.4161 + 0.2521i
  -0.0510 - 0.0073i  -0.0543 - 0.3239i   0.0481 + 0.3949i
   0.3684 - 0.0369i   0.3907 - 0.1531i  -0.2437 - 0.3094i
   0.2315 + 0.6895i   0.3656 - 0.1563i  -0.2134 - 0.1308i
   0.3656 - 0.1563i  -0.4318 + 0.6585i  -0.1541 + 0.0260i
  -0.2134 - 0.1308i  -0.1541 + 0.0260i  -0.4997 + 1.0021i.
\end{verbatim}
However, the square of the norm of the shortest lattice vector of the 
imaginary part of the matrix $W$ is $0.6585$, well below the threshold 
$\sqrt{3}/2$. This shows once more the limitations of the LLL 
algorithm since the convergence of the theta series we are interested 
in is controlled by the length of the shortest lattice vector. Note 
that the LLL reduced $\tilde{Y}$ above has the shortest vector in the 
6th column (with squared norm $0.6585$). One could construct a 
unimodular matrix $Z$ such that $T*Z$ has this vector appearing in the first 
column (the resulting matrix might not satisfy the LLL condition 
(\ref{LLLcond})). This would be more suited to the application of Siegel's 
algorithm, but will be still approximate since in general LLL does not identify 
the shortest lattice vector correctly. 

If the same algorithm is applied with an exact determination of the 
shortest vector, the picture changes considerably: in the first step 
of the iteration, the shortest lattice vector is correctly identified 
having the square of the norm  $0.6585$. Thus after a shift of the 
real part, a quasi-inversion is applied. The subsequent 
identification of the shortest vector of the resulting matrix leads 
to a vector of squared norm $0.7259$. After a shift of the real part, 
another quasi-inversion is applied. This time the square of the norm 
of the shortest vector is $1.0211$ and thus greater than 
$\sqrt{3}/2$. After a shift of the real part we finally obtain
\begin{verbatim}
W =

  Columns 1 through 4

   0.3967 + 1.0211i   0.0615 - 0.1322i  -0.0000 + 0.0000i  -0.4609 - 0.2609i
   0.0615 - 0.1322i   0.3967 + 1.0211i   0.3553 - 0.5828i  -0.3386 + 0.1933i
  -0.0000 + 0.0000i   0.3553 - 0.5828i   0.2894 + 1.1656i   0.0905 + 0.2450i
  -0.4609 - 0.2609i  -0.3386 + 0.1933i   0.0905 + 0.2450i   0.3967 + 1.0211i
   0.3553 - 0.5828i   0.4776 - 0.1287i  -0.4776 + 0.1287i  -0.4776 + 0.1287i
   0.1838 + 0.3219i   0.2743 + 0.5669i   0.3871 - 0.3736i   0.0167 - 0.3895i
  -0.3386 + 0.1933i  -0.3386 + 0.1933i  -0.1223 - 0.4541i   0.0615 - 0.1322i

  Columns 5 through 7

   0.3553 - 0.5828i   0.1838 + 0.3219i  -0.3386 + 0.1933i
   0.4776 - 0.1287i   0.2743 + 0.5669i  -0.3386 + 0.1933i
  -0.4776 + 0.1287i   0.3871 - 0.3736i  -0.1223 - 0.4541i
  -0.4776 + 0.1287i   0.0167 - 0.3895i   0.0615 - 0.1322i
   0.2894 + 1.1656i  -0.1671 - 0.7115i   0.0905 + 0.2450i
  -0.1671 - 0.7115i   0.4414 + 1.2784i  -0.3386 + 0.1933i
   0.0905 + 0.2450i  -0.3386 + 0.1933i   0.3967 + 1.0211i.
\end{verbatim}
In contrast to the algorithm incorporating LLL reductions, the squared length of
the shortest vector of the imaginary part is here given by the (11) component of
the matrix $W$. Note that the approximate character of the LLL algorithm is
unsatisfactory for our purposes for two reasons: First the overestimation of
the length of the shortest vector leads to a premature end of the algorithm
and a much shorter shortest vector than necessary. But secondly the potentially
crude approximation of its length implies that an estimate of the truncation
parameter $\mathcal{N}_{\epsilon}$ in (\ref{ne}) based on the LLL 
result could be misleading with the
consequence of a loss of accuracy in the approximation of the theta function.

Matlab timings have to be taken with a grain of salt since they 
depend crucially on the coding, in particular on how many precompiled 
commands could be used. Still in applications it is important to know 
how long a certain task takes on a given computer. For the above 
example, the LLL code is not very efficient, but converges in roughly 
1~ms. The SVP code takes in this case 4-5 times longer, which 
is still completely negligible compared to what can be gained by 
applying the above algorithm in the computation of a theta function 
associated to this surface.

The above example is in fact typical. If we consider an example of 
even higher genus, the curve
\begin{equation}
    f(x,y):=y^9+2x^2y^6+2x^4y^3+x^6+y^2=0
    \label{g16}
\end{equation}
of genus 16, we find a similar behavior. Using Siegel's algorithm on the Riemann
matrix for this curve computed with the code \cite{rsart}, we find that the
variant with the LLL algorithm converges within three iterations. The LLL
algorithm takes 1-2ms in each step. The algorithm produces
$\mathbb{B}_{11}=0.3314 + 1.0188i$, a value clearly larger than 1. The length of
the shortest vector generated by the imaginary part of this Riemann matrix as
found via SVP is $0.4437$, well below the theoretical minimum of
$\sqrt{3}/2\approx 0.866$.  On the other hand Siegel's algorithm with an exact
solution of the SVP in each step requires 14 iterations where each SVP takes
around 10ms. Finally we get $\mathbb{B}_{11}=0.4748 + 0.8956i$, i.e., a shortest
vector almost twice as long as what has been found with the LLL algorithm.

\section{Outlook}
In this paper, we have shown that for Siegel's algorithm (theorem 
\ref{thm}) can be used to efficiently compute multi-dimensional theta 
functions. For a genus $g>2$, an exact determination of the shortest 
vector of the lattice generated by the imaginary part of the Riemann 
matrix is recommended. The approximative LLL algorithm is for $g<20$ 
only an order of magnitude faster than the SVP algorithm, but 
finds the shortest vector merely with an error growing exponentially 
with~$g$. 

From the point of view of symplectic geometry, it would be 
interesting to find an algorithm to approach Siegel's fundamental 
domain (see definition \ref{def}) in a better way. This would allow to decide within 
numerical precision whether two different algebraic curves in fact 
define the same Riemann surface: a possible way to decide this would 
be to construct for both Riemann surfaces the symplectic 
transformations to the Siegel fundamental domain. If both Riemann 
matrices map to the same point in the fundamental domain, they 
correspond to the same surface. A problem in this context is that the 
Minkowski fundamental domain is only known for $g\leq 3$. Thus the 
case $g=3$ is the most promising to study in this context. Even less 
is known about the third condition in definition \ref{def}. It would 
be interesting to explore the matrices $C$ and $D$ there as 
Gottschling \cite{gottschling} did in genus 2 to compute 
$|\det(C\mathbb{B}+D)|$ at least for an interesting set of these 
matrices. The goal would be to approximate this maximal height 
condition better than with the quasi-inversion (\ref{quasi}). This 
will be the subject of further work. 

%

\noindent \textbf{Acknowledgement:}\\
We thanks D.~Stehl\'e for helpful discussions and hints.  
JF thanks for the hospitality at the University of Burgundy as a 
visiting professor, where part of this work has been completed.

\end{document}